\def\rn{\noindent\parshape 2 0truecm 16truecm 0.5truecm 15.5truecm}
\def\nn#1 #2{#1, #2.}				
\def\nnn#1 #2 #3{#1, #2. #3.}			
\def\nnnn#1 #2 #3 #4{#1, #2. #3. #4.}		
\def\nnnnn#1 #2 #3 #4 #5{#1, #2. #3. #4. #5.}	
\def\dualand{, \&\hbox{ }}				
\def\multiand{, \&\hbox{ }}				

\def\rg#1;#2;#3;#4;#5;#6 {\par\rn#1 #2, {\it #3}, {\bf #4}, #5 (``#6'') \par}
\def\rf#1;#2;#3;#4;#5 {\par\rn#1 #2, {\it #3}, {\bf #4}, #5\par}
\def\rfbook#1;#2;#3;#4;#5 {{\frenchspacing\par\rn#1 #2, {\it #3} (#4: #5)\par}}
\def\rfproc#1;#2;#3;#4;#5;#6 {{\frenchspacing\par\rn#1 #2, in {\it #3}, ed. #4 (#5: #6)\par}}
\def\rfprocp#1;#2;#3;#4;#5;#6;#7 {{\frenchspacing\par\rn#1 #2, in {\it #3}, ed. #4 (#5: #6), p#7\par}}
\def\rfprep#1;#2;#3  {{\par\rn#1 #2, #3\par}}
\def\rfprepp#1;#2;#3 {{\par\rn#1 #2, #3\par}}

\def\Mpc{{\rm Mpc}}

\def\kln{k_{\l n}}

\def\expec#1{\langle#1\rangle}

\def\etal{{\frenchspacing\it et al.}}
\def\ie{{\frenchspacing\it i.e.}}
\def\eg{{\frenchspacing\it e.g.}}

\def\beq#1{\begin{equation}\label{#1}}
\def\eeq{\end{equation}}
\def\beqa#1{\begin{eqnarray}\label{#1}}
\def\eeqa{\end{eqnarray}}
\def\eq#1{equation~(\ref{#1})}
\def\Eq#1{Equation~(\ref{#1})}
\def\eqn#1{~(\ref{#1})}

\newcommand{\beeq}{\begin{equation}} 
\newcommand{\beeqa}{\begin{eqnarray}}

\def\fig#1{Figure~\ref{#1}}
\def\Fig#1{Figure~\ref{#1}}

\def\sec#1{Section~\ref{#1}}

\def\spose#1{\hbox to 0pt{#1\hss}}
\def\simlt{\mathrel{\spose{\lower 3pt\hbox{$\mathchar"218$}}
     \raise 2.0pt\hbox{$\mathchar"13C$}}}
\def\simgt{\mathrel{\spose{\lower 3pt\hbox{$\mathchar"218$}}
     \raise 2.0pt\hbox{$\mathchar"13E$}}}
\def\simpropto{\mathrel{\spose{\lower 3pt\hbox{$\mathchar"218$}}
     \raise 2.0pt\hbox{$\propto$}}}

\def\ed{\end{document}}

\def\tr{\hbox{tr}\>}

\def\Mpc{{\rm Mpc}}

\def\nbar{{\bar n}}

\def\ith{i^{th}}

\def\f{{\bf f}}
\def\vk{{\bf k}}
\def\r{{\bf r}}
\def\vb{{\bf b}}
\def\k{{\bf k}}
\def\p{{\bf p}}
\def\phat{\widehat\p}
\def\ph{\widehat p}
\def\q{{\bf q}}
\def\r{{\bf r}}
\def\x{{\bf x}}
\def\vx{{\bf x}}
\def\y{{\bf y}}

\def\A{{\bf A}}
\def\B{{\bf B}}
\def\C{{\bf C}}

\def\F{{\bf F}}
\def\I{{\bf I}}

\def\PHI{{\bf\Psi}}
\def\PI{{\bf\Pi}}

\def\N{{\bf N}}
\def\P{{\bf P}}

\def\W{{\bf W}}

\def\S{{\bf S}}

\def\nmonte{N_{\rm monte}}

\def\vzero{{\bf 0}}

\def\psih{\widehat{\psi}}

\def\M{{\bf M}}

\def\rh{\widehat{\bf r}}

\def\tr{\hbox{tr}\>}
\def\dV{{d^3k\over (2\pi)^3}}
\def\dag{^\dagger}

\def\ngal{N_{\rm gal}}
\def\nx{{N_x}}
\def\ny{{N_y}}
\def\np{{N_p}}
\def\rmax{R_{\rm max}}
\def\mean{m}
\def\meanvec{\bf\mean}
\def\l{\ell}
\def\lmax{\l_{\rm max}}
\def\nmax{n_{\rm max}}
\def\Ylm{Y_{\l m}}

\def\ed{\end{document}}

\def\mz{m_{Zw}}

\def\tr{\hbox{tr}\,}

\def\ith{i^{th}}

\newlength{\tskip}

\newlength{\colwidth}
\newlength{\idlwidth}
\newlength{\smwidth}

\documentstyle[emulateapj,danonecolfloat]{article}

\begin{document}
\twocolumn[


\journalid{337}{15 January 1989}
\articleid{11}{14}

\submitted{Submitted to ApJ 1999 November 26, accepted 2000 October 20}

\title{The power spectrum of the CfA/SSRS UZC galaxy redshift survey}

\author 
  {Nikhil Padmanabhan\altaffilmark{1}$^,$\altaffilmark{3}
  Max Tegmark\altaffilmark{2}$^,$\altaffilmark{3}$^,$\altaffilmark{4}
  and 
  Andrew J. S. Hamilton\altaffilmark{5}\\
}  
\date{Submitted 1998, August 31}

\begin{abstract}
The combined CfA2/SSRS redshift catalog has 
recently been improved and made public.
We compute the redshift-space power spectrum of this catalog
using the regions at $-2.5^\circ\le\delta_{1950}\le 50^\circ$
north ($8^h\le\alpha_{1950}\le 17^h$) and
south ($20^h\le\alpha_{1950}\le 3^h$, $b\le -13^\circ$)
of the Galactic plane,
where it is 98\% complete down to Zwicky magnitude 15.5
and contains 13,681 galaxies.
Our analysis uses Heavens-Taylor mode expansion,
Karhunen-Lo\`eve data compression and 
the Fisher matrix technique to compute quadratic band power estimates.
This allows an exact calculation of window 
functions, including the integral constraint,
in addition to the production of a power spectrum
with uncorrelated error bars.
Our results with this larger data set agree well 
with previous studies. We analyze $101h^{-1}\Mpc$ and
$130 h^{-1}\Mpc$ volume-limited subsets in addition to the full magnitude-limited
sample, and our results are well fit by, \eg, 
$\Lambda$CDM models with bias $b=$1.2, 1.4 and 1.4, respectively.
We estimate the effect of extinction using the 
Schlegel, Finkbeiner \& Davis dust map.
Our results are exclusively for the redshift space
galaxy power spectrum, so they can only be compared with theoretical 
predictions if appropriate corrections are made for
biasing and redshift space distortions.
\end{abstract}


]


\altaffiltext{1}{Stanford University, Department of Physics, Stanford, CA
94305-4060; paddy@perseus.stanford.edu}

\altaffiltext{2}{University of Pennsylvania, Department of Physics, 
Philadelphia, PA 19104; max@physics.upenn.edu}

\altaffiltext{3}{Institute for Advanced Study, Princeton, NJ 08540}

\altaffiltext{4}{Hubble Fellow}

\altaffiltext{5}{JILA and Deptartment of Astrophysical 
and Planetary Sciences,
Box 440, Univ. of Colorado, Boulder, CO 80309; andrew.hamilton@colorado.edu}

\section{Introduction}
\label{IntroSec}

The three-dimensional maps of the Universe provided by
galaxy redshift surveys provide valuable information 
about many fundamental cosmological parameters.
This has motivated the creation of large 
data sets such as 
the Harvard-Smithsonian Center for Astrophysics (CfA),
Las Campanas (LCRS)
and IRAS 0.6 Jy (PSCz) redshift catalogs,
each well in excess of $10^4$ galaxies.
Even more ambitious projects are currently under way, 
with the AAT two degree field survey (2dF) aiming 
for 250,000 galaxies 
and the Sloan Digital Sky Survey
(SDSS) for 1 million. 
Combining such surveys with measurements from
upcoming CMB experiments,\eg, the MAP satellite, 
enables many cosmological parameters to be measured
much more accurately than would be possible using 
CMB information alone (Eisenstein {\etal} 1999) 
--- in principle.
Achieving this in practice requires that the matter 
power spectrum $P(k)$ can be accurately measured 
despite all the complications that are inevitably 
present in real-world redshift surveys.

This challenge has stimulated a large body of work 
over the last few years aimed at tackling such real-world 
issues, ranging from improved models of bias and extinction 
to the development of powerful new methods for 
measuring the power spectrum from surveys with arbitrary 
geometry and selection functions. 

The new CfA/SSRS UZC catalog has 
recently been completed (Falco {\etal} 1999), 
made public and had its correlation function (Girardi {\etal} 2000) and 
topology (Schmalzing \& Diaferio 2000) measured.
Since the power spectrum analyses of the
original CfA2 and SSRS surveys were performed some time ago
(Park {\etal} 1994;
da Costa {\etal} 1994; 
Marzke {\etal} 1994),
it is therefore quite timely to apply some of these new methods
to this extensive and further improved data set.
This is the purpose of the present work. 
We limit our analysis to the galaxy redshift space power spectrum on 
large scales, and therefore do not include the important complications 
of biasing, redshift space distortions or nonlinearities.

The rest of this paper is organized as follows. 
After describing the UZC data set in \sec{DataSec},
we present our analysis method in \sec{MethodSec}.
The results are given in \sec{ResultsSec}, which also tests their
robustness to various underlying assumptions. We summarize the conclusions
and remaining challenges in \sec{ConclusionSec}.

\section{Data}
\label{DataSec}

The previous CfA redshift surveys 
(Huchra, Vogeley \& Geller 1999; 
Huchra, Geller \& Corwin 1995;
Huchra {\etal} 1990; 
Geller \& Huchra 1989; 
Huchra {\etal} 1983;
Davis {\etal} 1982)
have all been based on the Zwicky catalog of galaxies 
with magnitude $\mz\le 15.5$, involving a  
heterogeneous sets of galaxy coordinates and redshifts.
Falco {\etal} (1999) recently completed 
and made public the Updated Zwicky Catalog (UZC), which was
improved over its predecessors in a number of ways:
\begin{enumerate}
\item Over 5,000 redshifts have been re-measured or uniformly re-reduced.
\item The accuracy of the galaxy coordinates has been improved to better than
two arcseconds using the digitized POSS plates.
\item The redshift ``blunder rate'' has been estimated and substantially reduced.
\end{enumerate}
The sample described in Falco {\etal} (1999) consists of 19285 galaxies, 
some without measured redshifts. The sample has subsequently been 
further improved and extended to 19415 galaxies, available online\footnote{
The latest version of UZC data set is available at\\ 
{\it cfa-www.harvard.edu/$\sim$falco/UZC}
}.

\begin{figure*}[tb] 
\vskip-4.5cm
\centerline{\epsfxsize=18cm\epsffile{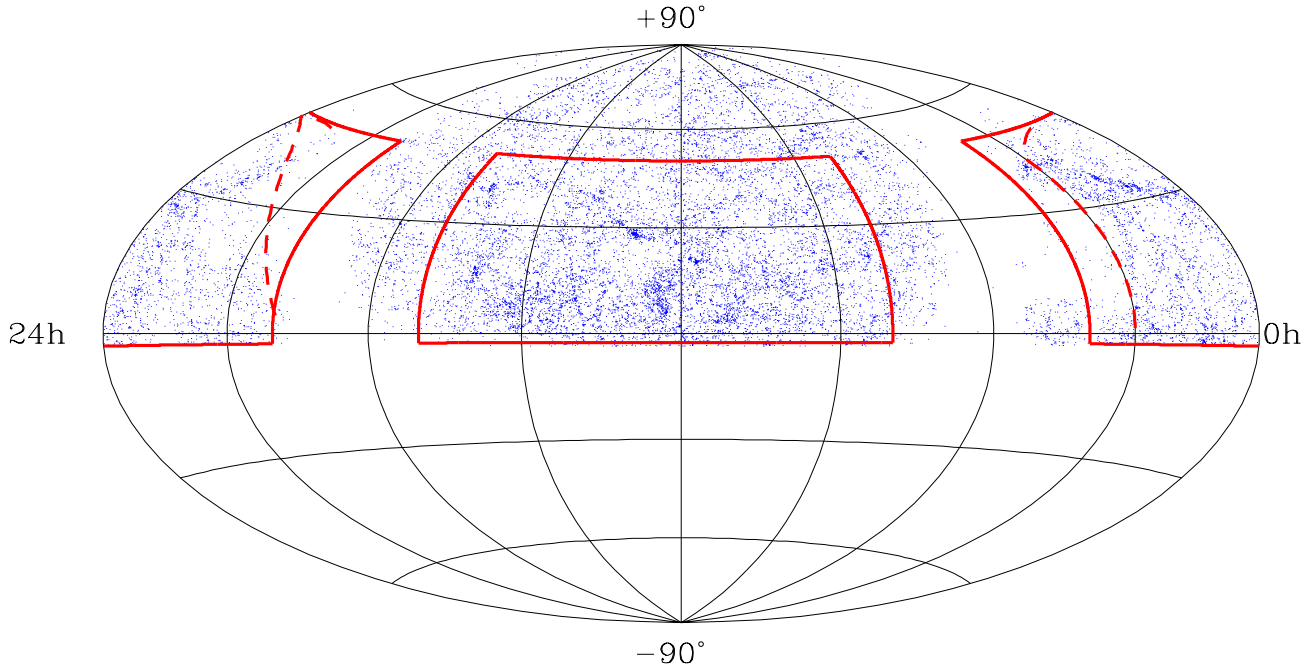}}
\vskip-4.5cm
\caption{\label{AitoffFig}\footnotesize%
The 19415 UZC galaxies are shown in Aitoff projection
in equatorial coordinates. The two 98\% complete subsets
are delimited by heavy lines, 
with the North Galactic subset in the center. We make additional 
cuts at $b=-13^\circ$ and $\alpha=3^h$ (dashed lines) 
to reduce extinction problems.
}
\end{figure*}

\begin{figure}[tb] 
\vskip-1.5cm
\centerline{\epsfxsize=9cm\epsffile{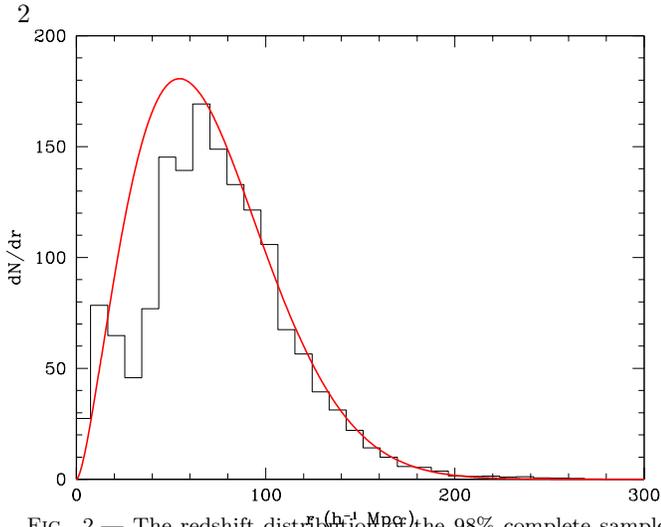}}
\vskip-1.5cm
\caption{\label{nbarFig}\footnotesize%
The redshift distribution of the 98\% complete sample of 13,681 galaxies.
The smooth curve is the selection function of de Lapparent et al 1989.
}
\end{figure}

We use the latest (as of November 1999) version of this data set, which consists
of 18,763 galaxies with redshifts.
Their angular distribution is shown in 
\fig{AitoffFig}. Our analysis is limited to the subset defined
by $20^h\le\alpha_{1950}\le 4^h$ (hereafter ``South'') and
$8^h\le\alpha_{1950}\le 17^h$ (hereafter ``North''),
$-2.5^\circ\le\delta_{1950}\le 50^\circ$,
where the sample is 98\% complete down to
Zwicky magnitude 15.5 (Falco {\etal} 1999).
We also apply a Galactic cut $|b_{II}|>13^\circ$ and discard
the southern Galactic region with $\alpha_{1950}>3^h$
to reduce extinction problems. 
This leaves a subset consisting of 13,681 galaxies
in a region subtending about a quarter of the sky
($10369$ square degrees $\approx 3.16$ steradians).
The redshift distribution of these galaxies is shown in
\fig{nbarFig}. 
Here and throughout this paper, we neglect the cosmic 
deceleration correction 
and define the distance to a galaxy in redshift space
as simply 
\beq{rDefEq}
r\equiv {cz\over H} = (3000 h^{-1}\,\Mpc)\>z,
\eeq
since $z\ll 1$, where $z$ is the galaxy redshift 
in the CMB rest frame.
\Fig{nbarFig} also shows the radial selection function that
we use in our analysis, taken from
de Lapparent {\etal} 1989.
It assumes a Schecter luminosity function
with parameters $M_*=-19.3 + 5 \log h$ and $\alpha=-1.1$.
We truncate this magnitude limit radially so that 
$10^{-3}h^{-1}\Mpc < r < 150 h^{-1}\Mpc$,
which leaves 13,184 galaxies.
The lower limit removes 32 galaxies with negative redshifts.
To facilitate comparison with prior work, 
we also analyze two volume-limited subsamples of the data,
where we keep only galaxies whose absolute magnitude is bright enough
that they would have been visible above the 
$\mz\le 15.5$ cut even if they were at the edge of the volume.
Following Park {\etal} 1994, we take these to 
have depth $101\>h^{-1}\,\Mpc$ and $130\>h^{-1}\,\Mpc$, which gives 
samples of size 2061 and 909, respectively, with a selection function
that should in principle be constant within the volume.

\section{Method}
\label{MethodSec}

We analyze the data using the methods described
in Tegmark {\etal} (1998, hereafter ``T98''). 
The analysis consists of the following four steps:
\begin{enumerate}
\item Heavens-Taylor pixelization
\item Karhunen-Lo\`eve compression
\item Quadratic band-power estimation
\item Fisher decorrelation
\end{enumerate}
We will now describe these steps in more detail.

\subsection{Step 1: Heavens-Taylor pixelization}

Our raw data consists of $\ngal$ three-dimensional
vectors $\r_\alpha$, $\alpha=1,...,\ngal$, giving the
measured positions of each galaxy in redshift space.
As discussed in T98, it is convenient to 
define the overdensity in $\nx$ 
``pixels'' $x_i$, $i=1,...,\nx$ by 
\beq{xDefEq}
x_i \equiv\int\left[{n(\r)\over\nbar(\r)}-1\right]\psi_i(\r) d^3r
= \int{n(\r)\over\nbar(\r)}\psi_i(\r) d^3r - m_i
\eeq
for some set of functions $\psi_i$
and work with the $\nx$-dimensional data vector $\x$ instead of the 
the $3\times\ngal$ numbers $\r_\alpha$.
Since the mean density term
\beq{meanDefEq}
\mean_i\equiv\int\psi_i(\r)d^3 r
\eeq
has been subtracted out, we have
\beqa{xExpecEq}
\expec{\vx}&=&\vzero,\\
\label{xCovEq}
\expec{\vx\vx^\dagger}&=&\C\equiv\N+\S,
\eeqa
where 
the shot noise covariance matrix given by
\beq{NdefEq}
\N_{ij} = \int {\psi_i(\r)\psi_j(\r)\over\nbar(\r)}d^3 r
\eeq
and the signal covariance matrix is 
\beq{SdefEq}
\S_{ij} = \int\psih_i(\vk)\psih_j(\vk)^*
P(k)\dV.
\eeq
Here hats denote Fourier transforms and
$\nbar$ is the three-dimensional selection function of the galaxy
survey, {\ie}, $\nbar(\r)dV$
is the expected (not the observed) number of galaxies in 
a volume $dV$ about $\r$.

\begin{figure}[tb] 
\centerline{\epsfxsize=9cm\epsffile{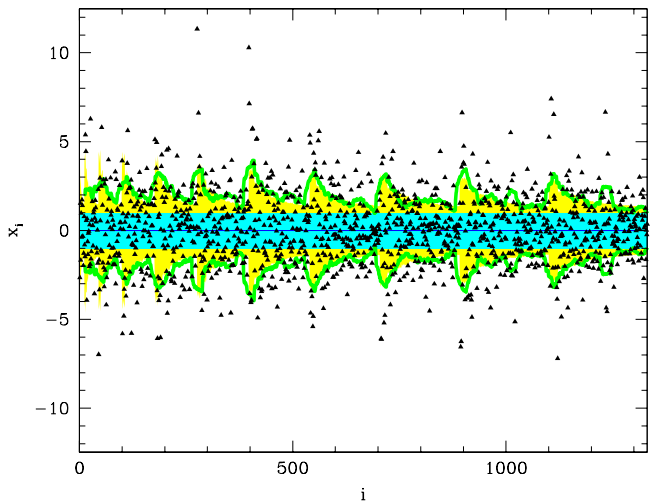}}
\caption{\label{xFig}\footnotesize%
The triangles show the 1331 elements $x_i$ of the data vector $\x$
(the HT expansion coefficients) 
for the $130\>h^{-1}\Mpc$ volume-limited sample.
If there were no clustering in the survey, merely shot noise, 
they would have unit variance, and about $68\%$ of them would be
expected to lie within the blue/dark grey band.
If our $\Lambda$CDM prior power spectrum were correct,
then the standard deviation would be larger, as indicated by the 
shaded yellow/light grey band. 
The green/grey curve is the rms of the data points
$x_i$, averaged in bands of width 25, and is seen to 
agree fairly well with the shaded band.
}
\end{figure}

We follow Heavens \& Taylor (1995, hereafter ``HT'') in choosing our 
functions $\psi_i$ to be spherical waves
\beq{HarmonicPixelEq}
\psi_i(\r) = w_i(r)\Ylm(\rh) j_\ell(\kln r),
\eeq
where $Y_{\ell m}$ is a spherical harmonic and $j_\ell$ a 
spherical Bessel function. This choice of functions
has the advantage (Fisher, Scharf, \& Lahav 1994; HT) 
of forming an almost complete set, 
separating large and small scales fairly well and 
allowing efficient computation of $\meanvec$ and $\C$.
The pixelized data vector $\x$ is shown in 
\fig{xFig} for one of the volume-limited subsamples.

Here and throughout, we use a single index $i$ to refer to the 
triplet $(\l m n)$ specifying an HT mode.
The wavenumbers $\kln$ are chosen as in HT so that 
the derivative of $j_\ell(\kln r)$ vanishes at 
a fixed radius $r=\rmax$ and 
$j_\ell(\kln r)$ has $n$ zeros before that.
We choose the weight functions $w_i$ to be of the 
form (Feldman, Kaiser \& Peacock 1994)
\beq{wDefEq}
w_i(r)\propto{\nbar(r)\over 1 + \nbar(r)P(\kln)},
\eeq
normalized such that the shot noise has unit 
variance, \ie, $\N_{ii}=1$.
Here $P$ is our prior guess for the power spectrum, 
which we will discuss in more detail 
in \sec{PriorSec}.
The results turn out to be rather insensitive to this 
weight function choice --- we will return to this issue 
in more detail in \sec{WeightCheckSec}.

We avoid the complex issues described in Tadros {\etal} (1999)
by using real-valued spherical harmonics, which
are obtained from the standard spherical harmonics by replacing $e^{im\phi}$
by $\sqrt 2\sin m\phi$, $1$, $\sqrt 2\cos m\phi$ for $m<0$, 
$m=0$, $m>0$
respectively.
HT show that apart from redshift-space distortions, 
the signal covariance matrix is given by
$\S = \PHI\P\PHI^t$,
where the diagonal matrix 
\beq{PmatrixDefEq}
\P_{ii'}=\P_{\l m n,\l' m' n'}\equiv\delta_{ii'} P(\kln)
\eeq
contains the effect of the power spectrum\footnote{In fact, 
\eq{PmatrixDefEq} is merely an approximation, valid 
when the survey depth $\ll\rmax$. We will discuss this issue in more
detail in \sec{HTgripeSec}.
} 
and the matrix 
\beq{PhiDefEq}
\PHI_{ii'}\equiv
\int 
w_{\l n}(r) 
j_\l(\kln r)\Ylm(\rh) 
j_{\l'}(k_{\l' n'} r)Y_{\l' m'}(\rh)
d^3 r
\eeq
which is $\psi_i(\r)$ expressed in $(\l m n)$-space and
contains the relevant information about the survey geometry.
When computing $\PHI$, we use the elegant time-saving
trick discovered by HT of expanding the
angular part of the selection function in spherical harmonics
and replacing the angular integral by a sum over Clebsch-Gordan 
coefficients.

\subsection{Step 2: Karhunen-Lo\`eve compression}

As can be seen in \fig{xFig}, most of the HT coefficients $x_i$ have a fairly 
low signal-to-noise ratio, \ie, $\S_{ii}/\N_{ii}\simlt 1$ and they
are dominated by shot noise. Moreover, the HT modes form a complete set 
over the entire sky, so they are overcomplete for our case of the
UZC catalog which covers only a fraction of the sky.
Both of these facts suggest that data compression may be possible, 
whereby almost all the cosmological information is retained in a
smaller set of modes. Such compression would accelerate
the matrix operations described  in step 3, 
where the number of operations required grows as the number of modes cubed.

\begin{figure}[tb] 
\centerline{\epsfxsize=9cm\epsffile{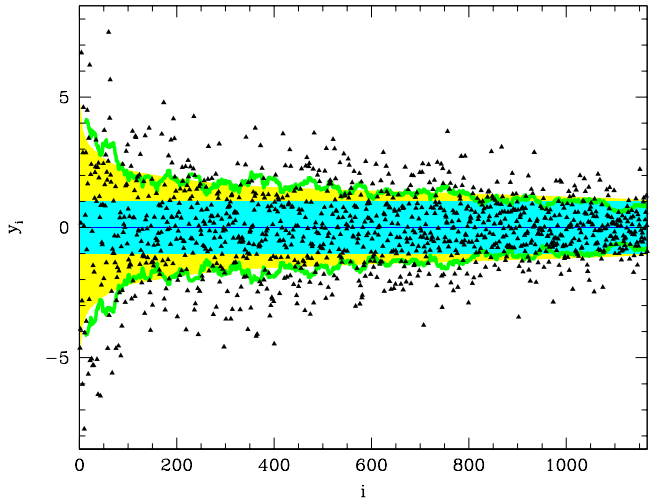}}
\caption{\label{WxFig}\footnotesize%
The triangles show the 1166 elements $y_i$ of the compressed data vector $\y$
(the KL expansion coefficients) 
for the $130\>h^{-1}\Mpc$ volume-limited sample.
If there were no clustering in the survey, merely shot noise, 
they would have unit variance, and about $68\%$ of them would be
expected to lie within the blue/dark grey band.
If our $\Lambda CDM$ prior power spectrum were correct,
then the standard deviation would be larger, as indicated by the 
shaded yellow/light grey band. The green/grey curve is the rms of the data points
$x_i$, averaged in bands of width 25, and is seen to 
agree better with the yellow/light grey band than the 
blue/dark grey band.
}
\end{figure}

We therefore subject the data vector $\x$ to Karhunen-Lo\`eve 
compression. This method (Karhunen 1947) 
was first introduced into large-scale structure analysis
by Vogeley \& Szalay (1996). It was
recently applied to the Las Campanas redsift survey
(Matsubara {\etal} 1999) and has been 
successfully applied to Cosmic Microwave Background data as well,
first by Bond (1995) and Bunn (1995).
We define a new data vector
\beq{zDefEq}
\y\equiv\B^t\x,
\eeq
where $\vb$, the columns of the matrix $\B$, are the $\nx$
eigenvectors of the generalized eigenvalue problem
\beq{SNeigenEq}
\S\vb = \lambda \N\vb,
\eeq
sorted from highest to lowest eigenvalue $\lambda$
and normalized so that $\vb\dag\N\vb=\I$.
This implies that
\beq{SNexpecEq}
\expec{y_i y_j}=\delta_{ij}(1+\lambda_i),
\eeq
which means that the transformed data values $\y$ have the desirable
property of being uncorrelated.
In the approximation that the distribution function
of $\x$ is a multivariate Gaussian, this 
also implies that they are statistically independent ---
then $\y$ is merely a vector of independent
Gaussian random variables. 
Moreover, \eq{SNeigenEq} shows that the eigenvalues $\lambda_i$ 
can be interpreted as a signal-to-noise ratio $S/N$.
Since the matrix $\B$ is invertible, the final data set $\y$ clearly retains
all the information that was present in $\x$.
In summary, the KL transformation partitions the information content of 
the original data set $\x$ into $\nx$ chunks that are
\begin{enumerate}
\item mutually exclusive (independent),
\item collectively exhaustive (jointly retaining all the information), and
\item sorted from best to worst in terms of their information content.
\end{enumerate}
\Fig{WxFig} shows that most of our KL coefficients $y_i$ have a signal-to-noise 
ratio $\lambda\ll 1$, so that the bulk of the 
cosmological information is retained in the first
$\ny$ coefficients, $\ny\ll\nx$.

In our case, numerical considerations force an additional compression
step. The above-mentioned overcompleteness means that 
certain linear combinations of modes vanish almost completely within
the angular mask of the survey (shown in \fig{AitoffFig}).
This makes many of of the eigenvalues of $\S$ and $\N$ 
essentially zero and poses a numerical problem when solving 
\eq{SNeigenEq}, since the standard reduction to an ordinary
eigenvalue problem by Cholesky decomposing $\N$ or $\S$ fails when 
both are singular.
Adding a small number to the diagonal elements of $\N$ also fails to
solve the problem: since these redundant modes are tiny,
avoiding to vanish completely mostly because of rounding errors,
they have a minute contribution from both signal and noise and therefore will
not automatically get weeded out by a cut on the signal-to-noise
$\lambda$. We therefore begin our compression 
by expanding the data in the eigenvectors of $\S$
and get rid of these redundant junk modes by throwing away all 
eigenmodes with eigenvalues below $10^{-6}$.
We then subject the new compressed data set to KL compression.

In conclusion, this step takes the vector $\x$ and its covariance matrix
$\C$ from \fig{xFig} and compresses it 
into the smaller vector $\B^t\x$ and its covariance matrix
$\B^t\C\B$, illustrated in \fig{WxFig}.

\subsection{Step 3: Quadratic band-power estimation}
\label{qSec}

\begin{figure}[tb] 
\centerline{\epsfxsize=9cm\epsffile{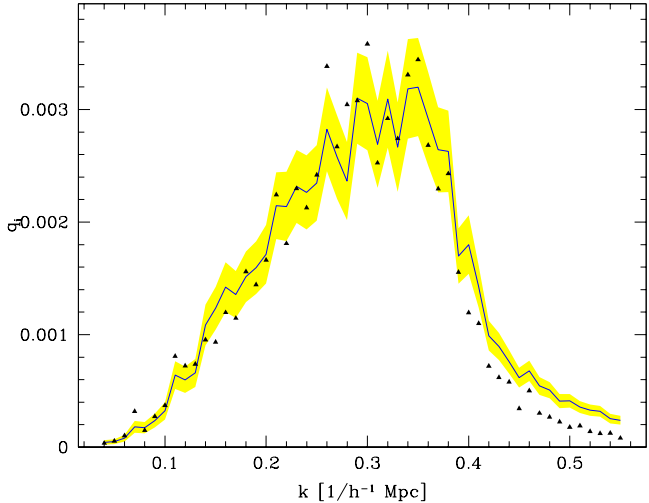}}
\caption{\label{qFig}\footnotesize%
The triangles show the 52 elements $q_i$ of the
raw quadratic band-power estimators $\q$
for the $130\>h^{-1}\Mpc$ volume-limited sample.
If the fiducial $\Lambda$CDM power spectrum were correct, 
then about $68\%$ of them would be
expected to lie within the shaded yellow/light grey band, 
centered on the solid curve.
}
\end{figure}

In this step, we perform a much more radical data compression
by taking certain quadratic combinations of the data vector
that can easily be converted into power spectrum measurements.

We parametrize the power spectrum as a piecewise constant function 
with $\np$ ``steps'' of height
$p_i$, which we term the {\bf band powers}
and group into an $\np$-dimensional vector $\p$.  
Thus $P(k) = p_i\ {\rm for}\ k_{i-1} \le k < k_i$, where
\beq{BandDefEq}
0=k_0<k_1<...<k_\np.
\eeq
We use $\np=52$ bands, most of them with bandwidth 
$\Delta k=0.01/h^{-1}\>\Mpc$, which should provide fine enough 
$k$-resolution to  
resolve any baryonic wiggles and other spectral features
that may be present in the power spectrum.
This means that we can write
\beq{CsumEq}
\C = \sum_{i=0}^{\np} p_i \C,{_i}
\eeq
where the derivative matrix 
$\C,{_i}\equiv\partial\C/\partial p_i$
is the contribution from the $\ith$ 
band and is computed by simply limiting the
implicit sums in the matrix multiplication 
$\S = \PHI\P\PHI^t$ 
to those lines and columns of the power spectrum matrix
$\P$ where $\kln$ lies in the $\ith$ band;
$k_{i-1}<\kln<k_i$.
For notational convenience, we included the noise term
in \eq{CsumEq} by defining $\C,{_0}\equiv\N$, corresponding
to an extra dummy parameter $p_0=1$ giving the shot noise normalization.

Our quadratic band power estimates are defined by
\beq{yDefEq2}
q_i\equiv {1\over 2}\x^t\C^{-1}\C_{,i}\C^{-1}\x,
\eeq
$i=1,...,\np$.
These numbers are shown in \fig{qFig}, and we will group them
together in an $\np$-dimensional vector $\q$.
Note that whereas $\x$ and $\y$ (and therefore $\C$ and $\B$) 
were dimensionless, $\p$ has units of power, \ie, volume.
\Eq{yDefEq2} therefore shows that
$\q$ has units of inverse power, \ie, inverse volume.
It is not immediately obvious that the vector $\q$ is a useful 
quantity. It is certainly not the final result (the power
spectrum estimates) that we want, since it does not even 
have the right units. Rather, like $\y$, it is a useful intermediate step.
In the approximation that the pixelized data has a 
Gaussian probability distribution (a good approximation in 
our case because of the central limit theorem, since $\ngal$ is large)
$\q$ has been shown
to retain all the information about the power
spectrum from the original data set (Tegmark 1997, hereafter ``T97'').
The numbers $q_i$ have the additional 
advantage (as compared with, \eg, maximum-likelihood estimators)
that their properties are easy to compute: their mean 
and covariance
are given by
\beqa{qExpecEq}
\expec{\q}&=&\F\p,\\
\label{qCovarEq}
\expec{\q\q^t}-\expec{\q}\expec{\q}^t &=&\F,
\eeqa
where $\F$ is the {\it Fisher information matrix} (Tegmark {\etal} 1997)
\beq{GaussFisherEq}
\F_{ij} = {1\over 2}\tr\left[\C^{-1}\C_{,i}\C^{-1}\C_{,j}\right].
\eeq
Quadratic estimators were first derived for galaxy survey
applications (Hamilton 1997ab). They were accelerated and 
first applied to CMB analysis (T97; Bond, Jaffe \& Knox 2000). 

In conclusion, this step takes the vector $\y$ and its covariance matrix
$\B^t\C\B$ from \fig{WxFig} and compresses it 
into the smaller vector $\q$ and its covariance matrix
$\F$, illustrated in \fig{qFig}.
Although \eq{qExpecEq} shows that we can obtain unbiased estimates of the 
true powers $\p$ by computing $\F^{-1}\q$, there are even better options,
as will be described in the next subsection. 

\subsection{Step 4: Fisher decorrelation}

Let us first eliminate the shot-noise dummy parameter $p_0$, 
since we know its value.
We define $\f$ to be the $0^{th}$ column 
of the Fisher matrix defined above
($f_i\equiv \F_{i0}$)
and restrict the indices $i$ and $j$ to run from
$1$ to $\np$ from now on, so 
$\f$, $\q$ and $\p$ are $\np$-dimensional vectors
and $\F$ is an $\np\times\np$ matrix.
Since $p_0=1$, \eq{qExpecEq} then becomes
$\expec{\q}=\F\p + \f$.

We now define a vector of shot noise corrected band power estimates
\beq{phatDefEq}
\phat\equiv\M(\q-\f),
\eeq
where 
$\M$ is some matrix whose rows are 
normalized so that the rows of
$\M\F$ sum to unity. 
Using equations\eqn{qExpecEq} and\eqn{qCovarEq}, this gives
\beqa{pExpecEq}
\expec{\phat}&=&\W\p,\\
\label{pCovarEq}
\expec{\phat\phat^t}-\expec{\phat}\expec{\phat}^t &=&\M\F\M^t,
\eeqa
where $\W\equiv\M\F$. We will refer to the rows of $\W$ as 
window functions,
since they sum to unity and \eq{pExpecEq} shows that
$\ph_i$ probes a weighted average of the true band powers $p_j$,
the $\ith$ row of $\W$ giving the weights.

\subsubsection{The minimum-variance choice}

\begin{figure}[tb] 
\vskip-1.0cm
\centerline{\epsfxsize=9cm\epsffile{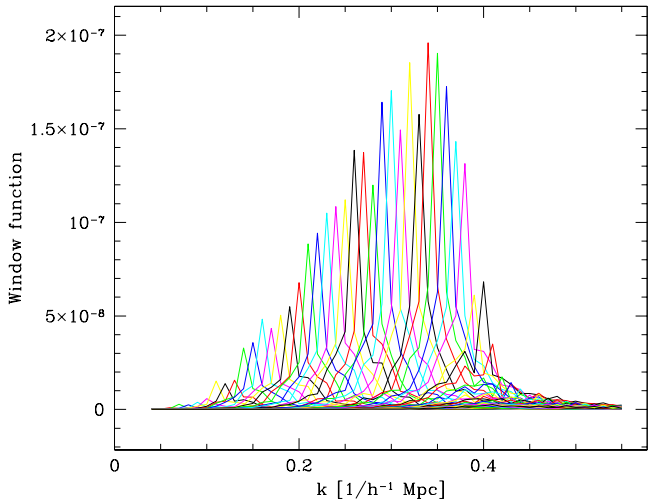}}
\vskip-1.0cm
\caption{\label{FisherFig}\footnotesize%
The rows of the Fisher matrix are shown
for the
$130\>h^{-1}\Mpc$ volume-limited sample.
The $\ith$ row typically peaks at the
$\ith$ band, the scale $k$ that the band power 
estimator $q_i$ was designed to probe.
Large amplitudes signify large information
--- if all curves had exactly the same shape, 
then the area under the $\ith$ curve would 
be $(\Delta\p_i)^{-2}$, the inverse variance of
$q_i$ when normalized as a band power estimator.
The turnover in the envelope at $k\sim 0.3/h^{-1}\Mpc$
reflects our omission of HT modes probing smaller scales.
}
\end{figure}

What is the best choice of the matrix $\M$?
A simple and natural choice is
\beq{SimpleChoiceEq}
\M_{ij} = \left(\sum_{j=1}^{\np}\F_{ij}\right)^{-1}\delta_{ij},
\eeq
\ie, $\M$ diagonal
(T97; Bond, Jaffe \& Knox 2000).
\Fig{FisherFig} plots the rows of the Fisher matrix, 
which have the same shape as the window functions for 
this choice of $\M$. 
As is seen in this figure, all elements of 
$\F$ are positive. 
This is guaranteed by the fact that the
matrices in \eq{GaussFisherEq} are all positive definite\footnote{
This can be seen as follows.
$\F_{ij} = {1\over 2}\tr[\A_i\A_j]$, where
$\A_i\equiv\C^{-1/2}\C_{,i}\C^{-1/2}$.
The trace of a such a product of two matrices is positive
if both matrices are positive definite --- this is obvious in 
the basis where one of them is diagonal, since neither can
have negative diagonal elements.
$\C_{,i}$ is positive definite since it is a covariance matrix 
(corresponding to a power spectrum equal to unity in 
the $\ith$ $k$-band and vanishing elsewhere), so the
matrices $\A_i$ are all positive definite.
}.

\subsubsection{The unbiased choice}

Another interesting choice is (T97)
\beq{UnbiasedChoiceEq}
\M=\F^{-1},
\eeq
which gives $\W=\I$.
In other words, all window functions are Kronecker
delta functions, and $\ph$ gives completely unbiased estimates 
of the band powers, with 
$\expec{\ph}=p_i$ regardless of what values the other band 
powers take.
A drawback of this choice is that the new covariance matrix
of \eq{pCovarEq} becomes $\F^{-1}$, which usually gives
substantially larger error bars ($\Delta p_i\equiv\M_{ii}^{1/2}=[(\F^{-1})_{ii}]^{1/2}$)
than the first method, anti-correlated between neighboring bands.

\subsubsection{The uncorrelated choice}
\label{UncorrSec}

\begin{figure}[tb] 
\vskip-1.0cm
\centerline{\epsfxsize=9cm\epsffile{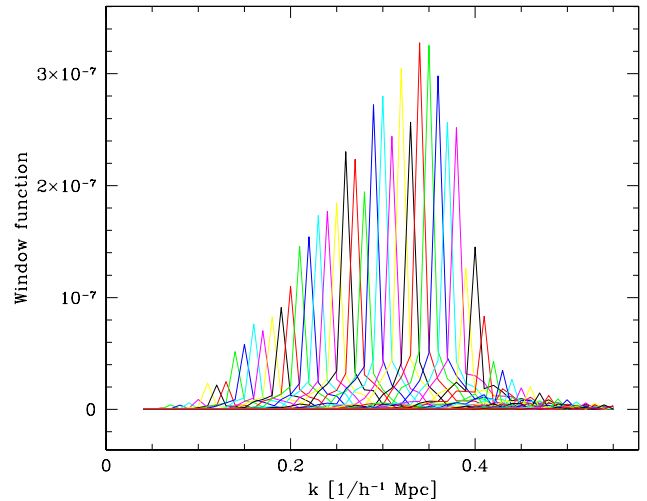}}
\vskip-1.0cm
\caption{\label{Wfig}\footnotesize%
The window functions are shown
for the
$130\>h^{-1}\Mpc$ volume-limited sample
using the decorrelation method.
The $\ith$ row of $\W$ typically peaks at the
$\ith$ band, the scale $k$ that the band power 
estimator $\ph_i$ was designed to probe.
To facilitate comparison with \fig{FisherFig}, we have
rescaled the window functions to sum to $(\Delta\p_i)^{-2}$
rather than unity here.
}
\end{figure}

The two above-mentioned choices for $\M$ both tend to 
produce correlations between the band power error bars.
The minimum-variance choice generally gives 
positive correlations, since the Fisher matrix cannot 
have negative elements, whereas 
the unbiased choice tends to give
anticorrelation between neighboring bands.
The choice (Tegmark \& Hamilton 1998; Hamilton \& Tegmark 2000) 
\beq{SqrtChoiceEq}
\M_{ij} = \left[\sum_{j=1}^{\np}\left(\F^{-1/2}\right)_{ij}\right]^{-1}\left(\F^{1/2}\right)_{ij},
\eeq
\ie, $\M=\F^{-1/2}$ with the rows renormalized,
has the attractive property of making the errors uncorrelated,
with the covariance matrix of \eq{pCovarEq} 
diagonal. The corresponding window functions $\W$ are
plotted in \fig{Wfig}, and are seen to be 
quite well-behaved, even narrower than those in
\fig{FisherFig} while remaining positive.
This choice is a compromise between the two first ones:
it narrows the minimum variance window functions at the cost of
only a small noise increase, with uncorrelated noise as an extra bonus.
Note that all three choices of $\M$ retain all the cosmological information,
since the vector $\y$ can always be recovered by multiplying $\phat$
by $\M^{-1}$.
The minimum-variance band power estimators are essentially a smoothed version 
of the uncorrelated ones, and 
their lower variance was 
paid for by correlations which reduced the effective number of independent 
measurements.

\subsection{Integral constraint correction}

Since the main focus of this paper is the power spectrum on 
the largest scales, it is crucial that we deal with the
complication known as the integral constraint (Peacock \& Nicholson 1991).
If we knew the selection function $\nbar(\r)$ {\it a priori}, 
before counting the galaxies in our survey, 
we would be able to measure 
the power on the scale of the survey. 
Our power spectrum estimate would essentially be the square of the ratio of 
the observed and expected number of galaxies in our sample.
Of course, we do not know $\nbar$ {\it a priori}, so we use the
galaxies themselves to normalize the selection function.
Thus the measured density fluctuation automatically 
vanishes on the scale of the survey, and 
and 
na\"\i ve 
application of the
power spectrum estimation method we have described
will falsely indicate that $P(k)\to 0$ as $k\to 0$, regardless of
the behavior of the true power spectrum on large scales.

Fortunately, this problem has a simple remedy once the data has been pixelized.
As described in T98,
the problem is that we do not know the true amplitude of the
mean density mode $\meanvec$ defined in \eq{meanDefEq}, and
consequently
may not have subtracted it out correctly in \eq{xDefEq}.
We can therefore immunize our data to this problem by making it orthogonal to $\meanvec$.
Defining a projection matrix
\beq{PIdefEq}
\PI\equiv\I-{\meanvec\meanvec^t\over|\meanvec|^2},
\eeq
the recipe is simply to replace $\x$ by $\PI\x$ and the matrices $\C,_i$ by
$\PI\C,_i\PI$ (T98). 
This trick of choosing modes that are orthogonal to
the mean density was first suggested by Fisher {\etal} (1993).
We find that this correction increases our error bars slightly on the 
largest scales, mainly in the leftmost band.

\section{Results}
\label{ResultsSec}

\subsection{Basic results}
\label{ResultsSubsec}

\begin{figure}[tb] 
\centerline{\epsfxsize=9cm\epsffile{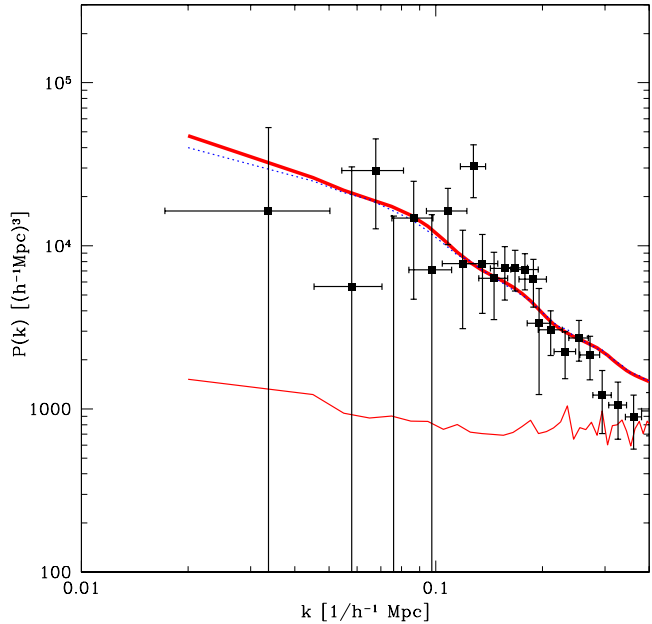}}
\caption{\label{101fig}\footnotesize%
Decorrelated band-power measurements are shown for the
$101\>h^{-1}\Mpc$ volume-limited sample. 
Although the window functions overlap slightly, the errors are 
all uncorrelated. The error bars include the
effects of both shot noise
and sample variance corresponding to the fiducial model. 
The horizontal bars show the rms
width of the window function corresponding to each measurement.
If the fiducial model (heavy curve) is correct, the measurements
should on average equal this curve convolved with the window functions
(dotted curve).
The thin red line shows the shot noise contribution that
has been subtracted out.
}
\end{figure}

\begin{figure}[tb] 
\centerline{\epsfxsize=9cm\epsffile{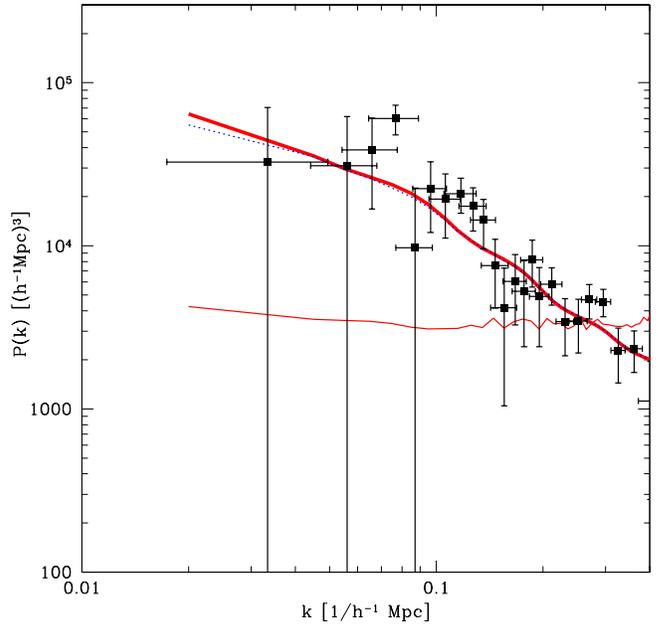}}
\caption{\label{130fig}\footnotesize%
Same as previous figure, but for the 
$130\>h^{-1}\Mpc$ volume-limited sample. 
}
\end{figure}

\begin{figure}[tb] 
\centerline{\epsfxsize=9cm\epsffile{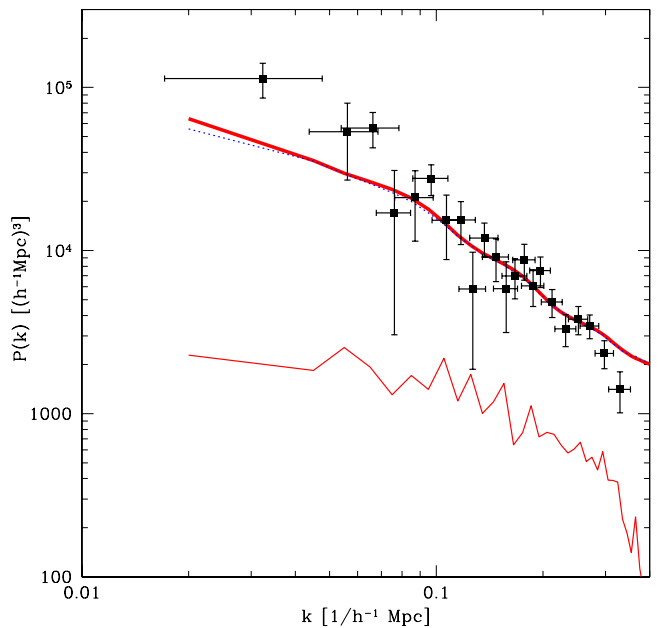}}
\caption{\label{magFig}\footnotesize%
Same as previous figure, but for the 
magnitude limited sample. 
}
\end{figure}

Our basic results are shown in Figures~\ref{101fig}, \ref{130fig}
and~\ref{magFig}.
Here the band power estimates are computed using the
$\M$-choice of \sec{UncorrSec}, so the error bars are uncorrelated.
The horizontal location of each data point and the width of the
horizontal bars are the mean and the standard deviation of the 
corresponding window function, respectively. To avoid excessive clutter,
we have averaged neighboring band-power measurements (and their 
corresponding window functions) on the smallest scales
using a simple inverse-variance weighting since the
error bars are uncorrelated.

The heavy curve shows the prior power spectrum vector $\p$
that was used in 
the calculation. 
It is a flat $\Lambda$CDM ``concordance model'' (Wang {\etal} 1999)
with $\Omega_\Lambda=0.7$, $\Omega_{cdm}=0.25$,
$\Omega_b=0.05$, $h=0.65$, $n=1$
with $\sigma=1$ for the matter fluctuations, rescaled by a different
bias factor for each of the three galaxy samples.
The linear power spectrum for this model was computed using
the fit of Eisenstein \& Hu (1999), then corrected for
nonlinear effects with the formalism of Jain {\etal} (1995) 
based on HKLM scaling 
(Hamilton {\etal} 1991; Peacock \& Dodds 1996), with the
local power spectrum slope given by the ``baryon wiggle-free'' 
version of the spectrum.
Our non-linear $\sigma_8=1$ normalization corresponds to
a linear $\sigma_8\approx 0.93$.
We first estimated a power spectrum assuming $b=1$, then
iterated the calculation once with new bias factors providing 
a better normalization to the actual measurements.
For the $101\,h^{-1}\Mpc$, $130\,h^{-1}\Mpc$
and magnitude-limited samples, these
three bias factors $b$ (reflected by the heavy curves 
shown in Figures~\ref{101fig}, \ref{130fig}
and~\ref{magFig}) are 1.2, 1.4 and 1.4, respectively.

This is not a complete treatment of non-linearity
(see, \eg, Meiksin \& White 1999; Scoccimarro {\etal} 1999; Hamilton 2000;
Benson {\etal} 1999).
The effects of nonlinearity will not bias our power
spectrum estimates, but the error bars are likely to 
be underestimated 
on nonlinear scales $k\simgt 0.3/h^{-1}\Mpc$.

Since each data point is the power spectrum convolved with a window
function, we would not expect the data points to fall exactly on
the true power spectrum even on average.
Rather, if the prior were correct, they would on average fall on 
the dotted curve $\W\p$, which shows
the prior averaged with the window functions of each band.
In all cases, the results are seen to be consistent with the 
priors used, except perhaps on the very smallest scales and
for the
leftmost band in the magnitude-limited case, to which we will
return in \sec{RadialErrorSec} below. 
We will discuss robustness towards the 
choice of prior below in \sec{PriorSec}.

How reliable are these results? 
In the remainder of this section, we present a series of tests, 
both of our software and algorithms
and of potential systematic errors.

\subsection{Validation of method and software}

Since our analysis consists of a number of somewhat complicated steps,
it is important to test both the software and the underlying methods.
We do this by generating $\nmonte=200$ Monte Carlo simulations of the
UZC catalog with a known power spectrum, processing them through our
analysis pipeline and checking whether they give the correct answer
on average and with a scatter corresponding to the predicted error bars.
We found this end-to-end testing to be quite useful in all phases of this 
project --- indeed, things worked on neither 
the first nor the second attempt...

\subsubsection{The mock survey generator}

Standard N-body simulations would not suffice for our precision test,
because of a slight catch-22 situation: the true non-linear power spectrum
of which an N-body simulation is a realization (with shot noise added)
is not known analytically, and is usually estimated by measuring it
from the simulation --- but this is precisely the step that we wish
to test.
We therefore resort to a simpler approach, where we generate 
realizations that are still in the linear regime, just as is
routinely done when preparing initial conditions for 
N-body simulations. We do this in the following steps:
\begin{enumerate}
\item Generate a Gaussian random field with the prescribed power spectrum
on a cubic grid (by generating uncorrelated Gaussian random variables with
variance $P(k)$
at each grid point in Fourier space, then performing an FFT).
\item Generate a random galaxy position $\r_i$ inside the angular mask
with radial probability distribution prescribed by the selection function.
\item Evaluate the density field $\delta$ at $\r_i$ using trilinear
interpolation between the nearest points on the FFT grid.
\item Add this galaxy to the catalog with a probability equal to
$[1+\delta(\r_i)]/2$, otherwise discard it.
\item Go back to step 2 and repeat until the desired number of galaxies
has been generated.
\end{enumerate}
To avoid the quantity $1+\delta(\r_i)$ going negative in step 4, which would
spoil our procedure, we normalize our fiducial power spectrum to have 
much less power than observed in the actual Universe. We choose our
test power spectrum to be a simple
Gaussian $P(k)\propto e^{-(Rk)^2/2}$ with
$R=32\,h^{-1}\Mpc$, normalized so that the rms fluctuations
$\expec{\delta^2}^{1/2}=0.2$.
This ensures that $|\delta|>1$ (breaking step 4) 
occurs only a negligible fraction 
of the time (about once in 1.7 million).

\subsubsection{Testing the Heavens-Taylor pixelization}

\begin{figure}[tb] 
\centerline{\epsfxsize=9cm\epsffile{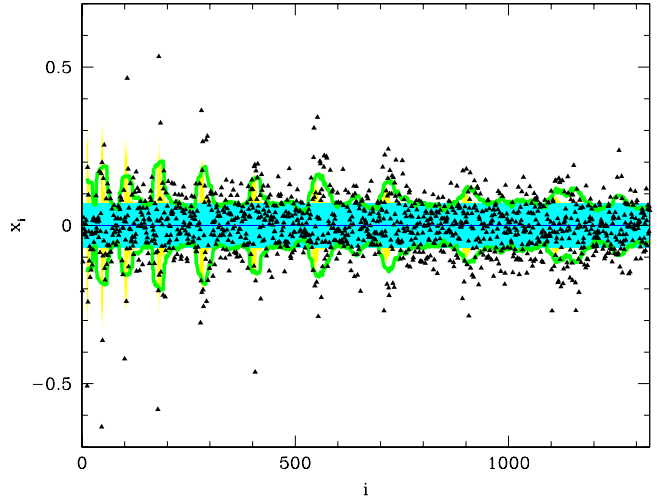}}
\caption{\label{xMonteFig}\footnotesize%
The triangles show the elements $x_i$ of the data vector $\x$
(the HT expansion coefficients) averaged
over 200 Monte-Carlo simulations of the 
$101\>h^{-1}\Mpc$ volume-limited sample.
If the algorithms and software are correct, then 
their mean should be zero and about
$68\%$ of them should lie within
the shaded yellow/grey region giving their standard deviation.
The green/grey curve is the rms of the data points
$x_i$, averaged in bands of width 25, and is seen to 
agree well with a smoothed version of the shaded region.
}
\end{figure}

\Fig{xMonteFig} shows the result of processing the Monte Carlo simulations
through the first step of the analysis pipeline, \ie, 
computing the corresponding Heavens-Taylor expansion coefficients
$x_i$. This is a very sensitive test of the mean correction 
given by \eq{meanDefEq}, which can be a couple of orders of magnitude
larger than the scatter in \fig{xMonteFig} for some modes.  
A number of problems with the 
radial selection function integration and the spherical harmonic expansion
of the angular mask in our code were discovered in this way.
After fixing these problems, the coefficients $x_i$ became consistent
with having zero mean as seen in the figure.
The figure also shows that the scatter in the modes
is consistent with the predicted standard deviation 
$\sigma_i=(\C_{ii}/\nmonte)^{1/2}$ (shaded region), with most of the
the fluctuations being localized to modes probing large scales
(with $\ell$, $m$ and $n$ being small).

%
%
%
Processing the Monte Carlo simulations
through the second step of the analysis pipeline
showed that
the corresponding KL eigenmodes $y_i$ passed the same test.


\subsubsection{Testing the quadratic band-power estimation}

\begin{figure}[tb] 
\centerline{\epsfxsize=9cm\epsffile{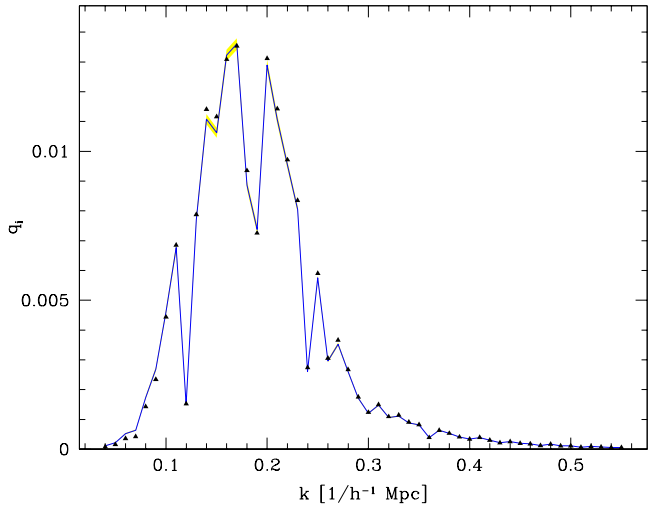}}
\caption{\label{qMonteFig}\footnotesize%
The triangles show the elements $q_i$ of the
raw quadratic band-power estimators $\q$, averaged
over 200 Monte-Carlo simulations of the 
$101\>h^{-1}\Mpc$ volume-limited sample.
If the algorithm and the software is correct, then 
about $68\%$ of them would be
expected to lie within the shaded yellow/grey band, 
centered on the solid curve.
}
\end{figure}

\Fig{qMonteFig} shows
the result of processing the Monte Carlo simulations
through the third step of the analysis pipeline, \ie, 
computing the raw unnormalized quadratic band-power estimates $q_i$. 
Since information from large numbers of modes contributes to each 
$q_i$, the scatter $\sigma_i=(\F_{ii}/\nmonte)^{1/2}$
is seen to be small. Therefore,
even quite subtle bugs and inaccuracies can be (and were!)
discovered and remedied as a result of this test.

\subsubsection{Testing the Fisher decorrelation}

\begin{figure}[tb] 
\centerline{\epsfxsize=9cm\epsffile{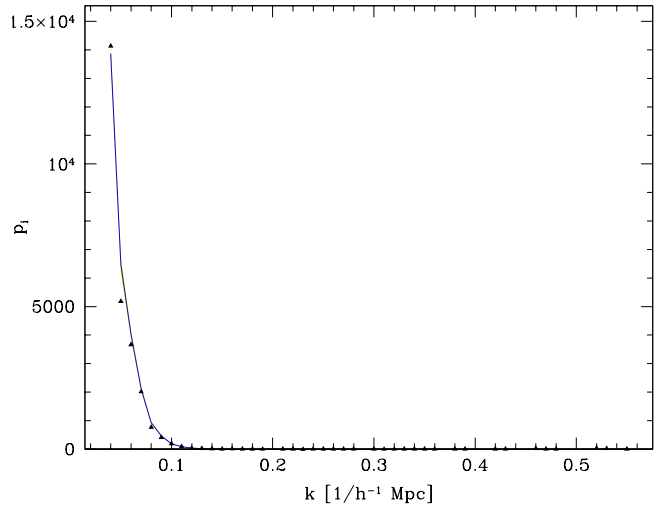}}
\caption{\label{pMonteFig}\footnotesize%
The triangles show the decorrelated band-power estimates $\ph_i$, 
averaged
over 200 Monte-Carlo simulations of the 
magnitude-limited 
sample.
If the algorithm and the software is correct, then 
about $68\%$ of them should 
lie within the shaded yellow/grey band, 
centered on the Gaussian fiducial power spectrum (solid curve).
}
\end{figure}

\begin{figure}[tb] 
\centerline{\epsfxsize=9cm\epsffile{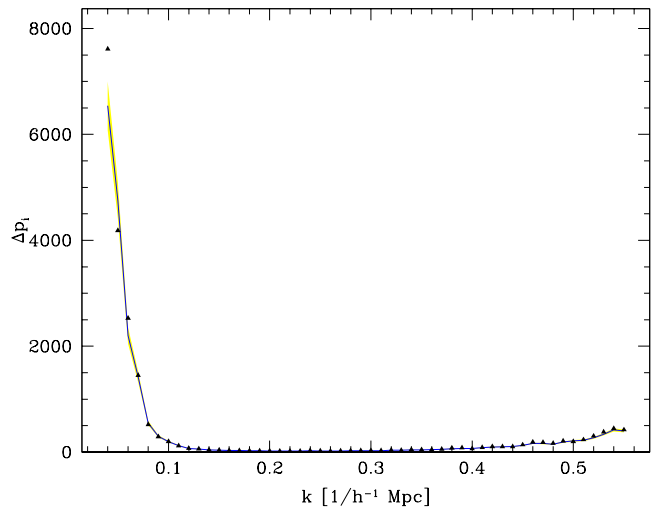}}
\caption{\label{pMonteFig2}\footnotesize%
Same as the previous figure, but for the error bars $\Delta p_i$.
The triangles show the observed scatter in the 200 
simulations. If the algorithm and the software is correct, then 
about $68\%$ of them should 
lie within the shaded yellow/grey band predicted by the Fisher matrix
formalism,
centered on the solid curve.
}
\end{figure}

Figures~\ref{pMonteFig} and~\ref{pMonteFig2} show 
the result of processing the Monte Carlo simulations
through the fourth and final step of the analysis pipeline, \ie, 
computing the decorrelated and normalized 
band-power estimates $p_i$. 
The mean recovered power spectrum is seen to be in excellent agreement
with the Gaussian prior used in the simulations (\fig{pMonteFig})
convolved with the window functions,
and the observed scatter is seen to be consistent
with the predicted error bars (\fig{pMonteFig2}).
These two figures therefore constitute an end-to-end test
of our data analysis pipeline, since errors in any of the
many intermediate steps would have shown up here at some level.

\subsection{Robustness to method details}

Our analysis pipeline has a number of ``knobs'' that can be
set in more than one way. This section discusses the 
sensitivity to such settings.

\subsubsection{Effect of changing the prior}

\label{PriorSec}

\begin{figure}[tb] 
\centerline{\epsfxsize=9cm\epsffile{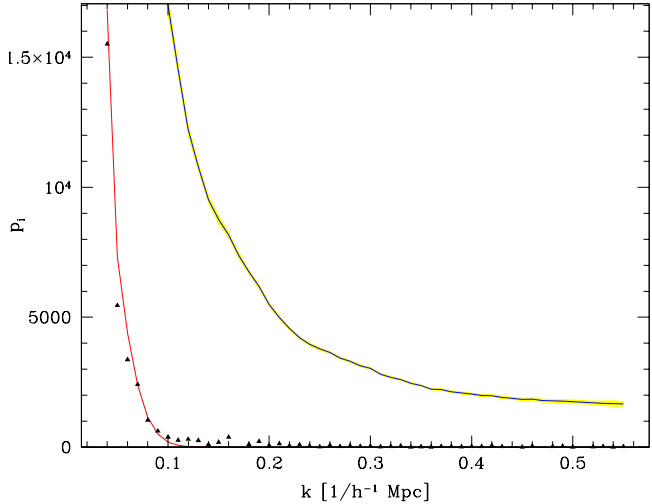}}
\caption{\label{GorskiFig}\footnotesize%
Effect of prior. The 
triangles show the decorrelated band-power estimates $p_i$,
averaged over 200 Monte-Carlo simulations of the 
magnitude-limited 
sample with a Gaussian power spectrum (red line).
The power spectrum estimation was carried out 
assuming a totally different fiducial model, 
the $\Lambda$CDM model to the right.
If the data was consistent with the prior, 
about 68\% of the measurements should 
fall within the narrow yellow/grey region. 
Instead, the method is seen to faithfully
reproduce the actual input power spectrum, with no evidence
of a bias towards the assumed prior.
}
\end{figure}

The analysis method employed assumes a ``prior'' power
spectrum via \eq{PmatrixDefEq}, both to compute band power error bars
and to find the galaxy pair weighting that minimizes them.
As mentioned, an iterative approach was adopted starting
with a simple $\Lambda$CDM model with $\sigma_8=1$, then 
rescaling it to better fit the resulting measurements and 
recomputing the measurements a second time.
To what extent does this choice of prior affect the results?
On purely theoretical grounds
(\eg, Tegmark, Taylor \& Heavens 1997), one expects a grossly incorrect
prior to give unbiased results but with unnecessarily large
variance. If the prior is too high, the sample-variance 
contribution to error bars will be overestimated and vice versa.
This hypothesis has been extensively 
tested and confirmed in the context of power spectrum measurements
from the Cosmic Microwave Background 
(\eg, Bunn 1995). An analogous test is shown in \fig{GorskiFig},
showing that the correct result is recovered even when our 
200 simulations are analyzed with a grossly 
incorrect prior. 

Generally, the pair weighting strives to minimize the joint 
contribution from sample variance and shot noise to the scatter in the 
measurements. This scatter will therefore be unnecessarily large 
both if the prior is too low (so that 
sample variance not taken seriously enough)
and if it is too high (so that excessive paranoia about sample variance 
gives a pair weighting producing unnecessarily large shot noise).
The resulting scatter therefore increases only to second order 
when the prior is slightly off.
Since the iterated priors used in our analysis
of the real data agree so well with the actual measurements,
slight remaining deviations of the prior from the truth 
are therefore likely to have a negligible impact.

\begin{figure}[tb] 
\centerline{\epsfxsize=9cm\epsffile{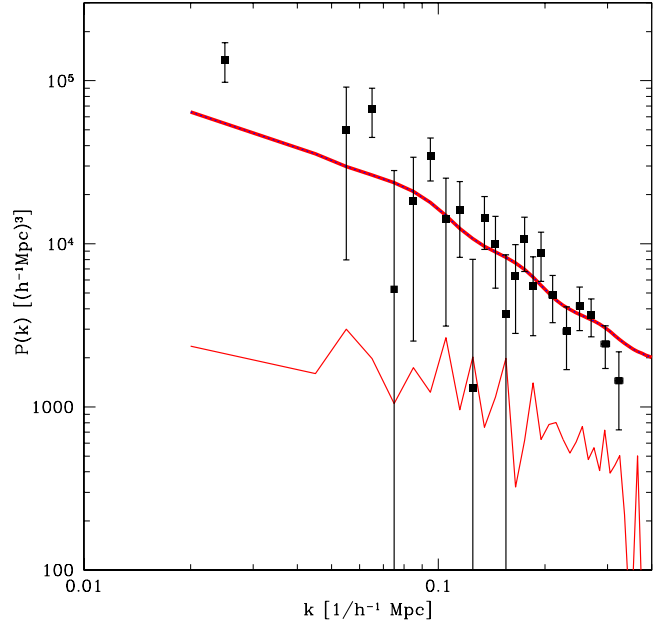}}
\caption{\label{AnticorrMagFig}\footnotesize%
Same as \fig{magFig} (magnitude-limited sample), 
but for the method giving maximally narrow window functions.
}
\end{figure}

\begin{figure}[tb] 
\centerline{\epsfxsize=9cm\epsffile{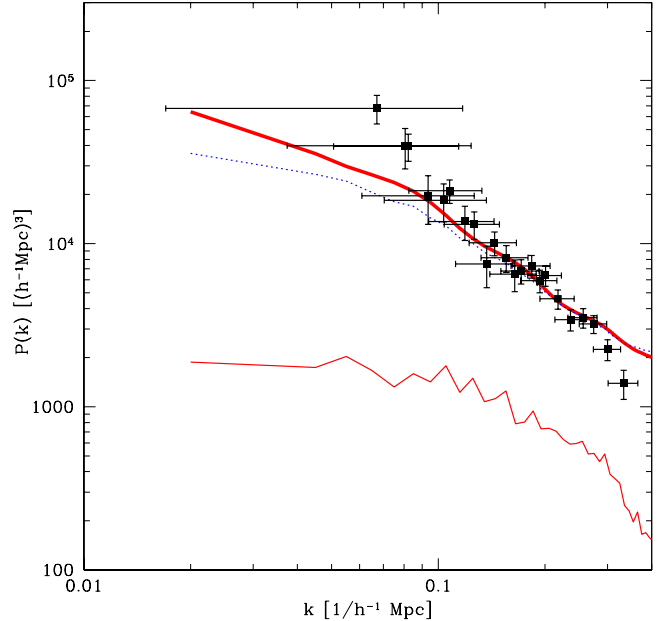}}
\caption{\label{CorrMagFig}\footnotesize%
Same as \fig{magFig} (magnitude-limited sample), 
but for the method giving smaller, bandwidth independent error bars.
}
\end{figure}

\subsubsection{Effect of changing the decorrelation method}

Our main results for the power spectra (Figures~\ref{101fig}--\ref{magFig})
were computed using the
decorrelation method given by \eq{SqrtChoiceEq}.
To assess the sensitivity to this choice, we repeated
the analysis for the other methods.
The results were generally consistent with the same 
fiducial models, but as expected, the nature of the scatter 
was found to be strongly method-dependent.
This is illustrated in 
Figures~\ref{AnticorrMagFig}
and~\ref{CorrMagFig} for the magnitude-limited case.

\Fig{AnticorrMagFig} used the method given by \eq{UnbiasedChoiceEq},
giving maximally narrow window functions. Although they are
plotted as having zero width, a calculation with narrower
bands would show them to have a width of order that of the bands used
here, \ie, of order the horizontal separation between neighboring 
points on the plot.
However, \fig{AnticorrMagFig} also shows that there 
are no free lunches: the errors are bigger than
in \fig{magFig}, and the error covariance matrix shows that they are 
strongly anti-correlated with their nearest neighbors.
This figure is essentially a window-deconvolved version 
of \fig{magFig}, and smoothing it would recover that figure.

\Fig{CorrMagFig} used the method given by \eq{SimpleChoiceEq},
and deviates from \fig{magFig} in the opposite way from the previous
example. It is essentially a smoothed version of \fig{magFig},
giving nice small error bars, slightly broader 
window functions and positively correlated errors between 
neighboring points. The broader window functions are seen to be 
particularly annoying on the largest scales, where they 
shift the effective wavenumber probed far to the right.

\subsubsection{Effect of changing the galaxy weighting}

\label{WeightCheckSec}

When expanding our galaxies in HT modes, we 
applied the radial FKP weighting given by 
\eq{wDefEq}. How does this particular choice affect the results?
To address this issue, we repeated the analysis with
\eq{wDefEq} replaced by $w(r)\propto\nbar(r)$, \ie, the simple
$P(k)=0$ limit of equal weight per galaxy. This resulted in 
almost no perceptible loss of information, typically increasing the
band-power error bars on large scales by less than a percent.
The resulting power spectrum measurements were essentially unchanged.

In the limit where infinitely many HT modes would be
used, any functions whatsoever can be created by taking linear
combinations of them, since they form a complete basis over the
survey volume. This would make the choice of the radial weighting 
function $w(r)$ completely irrelevant, since the subsequent 
KL compression step would end up recovering the true KL eigenfunctions
regardless. The reason that the radial weighting makes any difference
at all in our case is therefore that we have
used only a limited number of modes to start with, making it important
that they do not grossly over- or underweight sparse distant galaxies
relative to nearby ones.

\subsubsection{Effect of other method details}

\label{HTgripeSec}


The analysis above was performed using $11^3 = 1331$ HT modes, with
an angular cutoff at $\lmax=10$ giving 
$(1+\lmax)^2=121$ angular modes $\Ylm$ and a radial cutoff
$\nmax=10$ giving 11 radial modes $j_\l(k_{\l n} r)$ per $\Ylm$.
To explore the sensitivity to these choices, we 
repeated the entire analysis with $\lmax=\nmax=$~3, 5 and 7 giving 
64, 216 and 512 HT modes, respectively.
As expected, we found that the band powers on the very largest scales
converged quite rapidly as more modes were added, 
and that the new information made a difference
mainly on the smaller scales where the new modes were sensitive.
These numerical experiments suggested that our
1331 mode analysis retained a large fraction of the cosmological 
information down to $k\sim 0.3$, and that a more ambitious 
analysis with more modes would give substantially smaller
error bars at smaller scales.
The information matrix plotted in \fig{FisherFig} reenforces this conclusion.

To determine the best tradeoff between angular and radial modes,
we performed a number of tests with $\lmax\ne\nmax$, keeping
the total number of modes $(1+\lmax)^2(1+\nmax)$ roughly constant. 
These tests indicated that the largest Fisher 
information (the smallest error bars) on the power spectrum on
the largest scales 
was obtained for the symmetric case $\lmax\sim\nmax$. This is why 
we used $\lmax=\nmax$ in our main analysis.

\begin{figure}[tb] 
\vskip-1.0cm
\centerline{\epsfxsize=9cm\epsffile{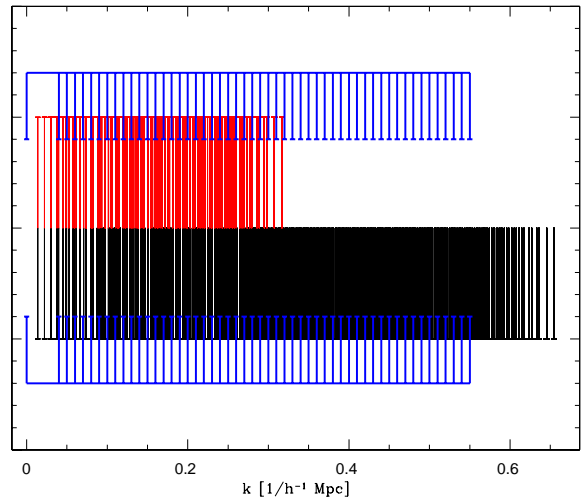}}
\vskip-1.0cm
\caption{\label{BandFig}\footnotesize%
Our 52 k-bands are shown (top and bottom) 
together with the discrete spectra of wavenumbers
$k_{\l m}$ used in the HT formalism.
Of the two spectra, the upper one corresponds to
the values $k_{\l m}$ used in \eq{HarmonicPixelEq} for
expanding the data, whereas the lower one 
corresponds to the values used in 
\eq{PmatrixDefEq} for computing the covariance matrix
$\C$ and its derivatives $\C,_i$, both for the magnitude limited 
analysis.
}
\end{figure}

We used 52 $k$-bands as illustrated in \fig{BandFig}.
To compute the covariance matrix $\C$ and its 52 derivatives 
$\C,_i$ exactly using \eq{PmatrixDefEq}, 
the contributions from a very large set of
$k_{\l m}$-values would be required. In practice,
we truncated the $k_{\l m}$-values used for these 
internal computations (the lower subset in \fig{BandFig})
at $\l=\lmax=21$, since experimentation with different cutoff values 
$\lmax$ showed that all results for $k\simlt 0.3$ had
converged by then. An alternative approach to truncation 
is to make a sharp cut at a fixed $k$-value (HT).

As mentioned above, \eq{PmatrixDefEq} is only approximate.
The discrete spectra shown in \fig{BandFig} would apply only 
if the radial Bessel functions were allowed to extend to infinite
radius. Since they are truncated outside the survey volume, 
the sharp lines of \fig{BandFig} get slightly smeared
out: they essentially get convolved with the Fourier transform of the
survey volume, which gives them a characteristic width $\Delta k$
of order the inverse size of the survey volume.

In the eigenmode expansion, the cutoff was placed at an $\S$-eigenvalue of
$10^{-6}$, simply to avoid numerical singularities. This reduced the original 1331
modes to 1182, 1166 and 1206 for the $101h^{-1}\Mpc$, $130h^{-1}\Mpc$ 
and magnitude limited samples, respectively.
The results remained completely unchanged when this 
$10^{-6}$ cutoff was increased by 
two orders of magnitude. Since 1331 models are very manageable numerically, 
our principal motivation for performing the subsequent S/N-analysis was 
to check for systematic errors. We therefore discarded no further modes in
this step.

\subsection{Robustness to extinction}

Mis-estimates of $\nbar$ constitute a 
potential source of systematic errors. Although our method for 
dealing with the integral constraint immunized against errors in 
the overall normalization of $\nbar$, errors in its shape would 
still add spurious power to our estimate of $P(k)$.
Let us first consider errors in the angular part of $\nbar$ caused by
Galactic dust, returning
to errors in the radial part in the next subsection.

The angular modulations caused by dust extinction tend to have a
power spectrum rising sharply towards the largest scales
(Vogeley 1998), and is therefore of particular concern for the interpretation
of our leftmost bandpower estimates. Although the UZC subset we are using is 98\% complete
relative to the underlying Zwicky sample, extinction can of course cause 
completeness modulations in the latter.
To estimate the severity of this effect, we used the
recent extinction map produced by Schlegel, Finkbeiner \& Davis
(1999), using their B-band conversion factor of 4.325.

\begin{figure}[tb] 
\vskip-1.5cm
\centerline{\epsfxsize=9cm\epsffile{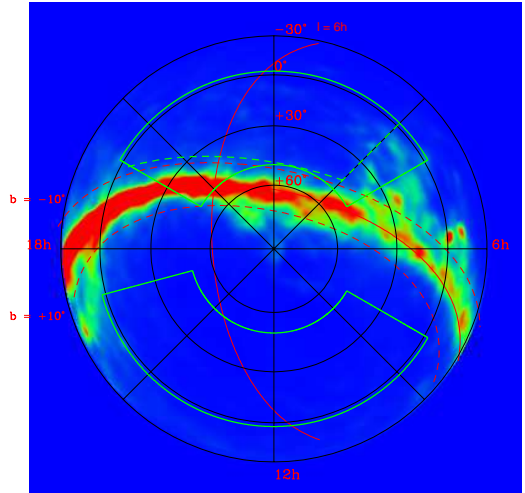}}
\vskip-3.4cm
\caption{\label{DustFig}\footnotesize%
The extinction $\Delta m$ predicted by Schlegel, Finkbeiner \& Davis (1999)
is shown in gnomonic equal-area 
projection with the North Celestial Pole in the
center and right ascension $\alpha=0$ at the top, increasing clockwise.
The Galactic plane is seen to intersect parts of the southern 
survey region. Note that the South Galactic Hemisphere is at the top.
}
\end{figure}

\begin{figure}[tb] 
\centerline{\epsfxsize=9cm\epsffile{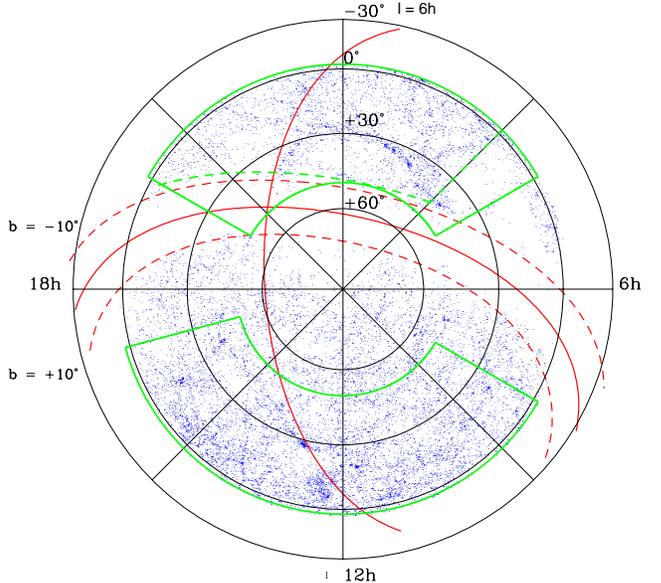}}
\vskip-0.5cm
\caption{\label{GnomoFig}\footnotesize%
The UZC galaxies are shown in the same coordinates as \fig{DustFig}.
Note the visible deficit of galaxies in regions near
the Galactic plane where Schlegel et al (1999)
predict high extinction.
}
\end{figure}

The expected extinction in the regions relevant to the 
UZC survey is shown in \fig{DustFig}, and the data set is 
shown in the same projection in \fig{GnomoFig} for comparison.
In the northern sample, the extinction ranges from 
$\Delta m=$0.012 to 1.7 with a median of 0.11.
The cleanest spot in the southern sample has $\Delta m=$0.061,
and the extinction gets as extreme as $\Delta m=$63 in
the Galactic plane (which we masked out).

To get a first crude handle on the importance of extinction, 
we applied our analysis separately to the relatively clean 
north and to the full, uncropped south (for this test, 
we used the regions delimited by
solid lines in \fig{GnomoFig}, not the dashed lines). 
Perhaps surprisingly, the South does not show a great
power excess over the North even though the whopping extinction 
associated with the Galactic plane was included in this southern sample.
This is presumably due to a combination of effects:
The calculations of Vogeley (1998) suggested that extinction (for 
the SDSS region in the north) would dominate over the cosmological power
spectrum only on scales $k\simlt 0.03/h^{-1}\Mpc$, and the smaller volume of
the UZC survey precludes us from effectively probing such large scales.
Also, although extinction is severe near the Galactic plane, 
this is a relatively smooth feature and therefore does not greatly 
affect smaller scale fluctuations. 

Indeed, the power spectrum in the North is, if anything,  
somewhat greater than in the South. This agrees
with the findings of Park {\etal} and da Costa {\etal} (1994), 
reflecting the
fact that the nearby southern sky is more quiescent, 
lacking northern structures such as Virgo and the Great Wall.

To be cautious, we nonetheless subjected the southern 
subsample to two additional cuts based on the dust map in \fig{DustFig}
before producing our main results, the power spectra shown in
\sec{ResultsSubsec}.  
Following Park {\etal} (1994), we 
moved the $4^h$ right ascension cut to $3^h$. We also excluded the
region less than $13^\circ$ from the Galactic plane. Since 
the north-south differences were relatively minor without these cuts, 
we expect extinction to play only a subdominant role in this cropped
data set.

We also corrected the observed magnitudes with the 
Schlegel {\etal} map and considered creating a new magnitude-limited sample 
for analysis. Unfortunately, this would have reduced the galaxy count 
too drastically to be of much interest. Since the
extinction gets as high as $\Delta m=1.7$ even in the north, 
the new magnitude 
cut would have to be shifted from 15.5 to 13.8 to ensure completeness, 
leaving only $8\%$ of the galaxies from the original sample.

\subsection{Robustness to radial selection function errors}
\label{RadialErrorSec}

To assess the extent to which radial selection function errors
may be adding spurious power, we repeated our analysis with
a variety of different selection functions. 
We replaced our Schecter parameters $(-1.1,-19.3)$ by a grid in 
$(\alpha,M_*)$-space, always normalizing to match the observed number of
galaxies. This had a substantial effect on the measured band powers, especially
on the largest scales. We experimented with an 
iterative approach whereby the selection function was
fine-tuned to minimize the large-scale power, but this unfortunately failed 
to converge: by grossly over-estimating the (small) number of distant 
galaxies, this procedure causes $\delta\equiv(n/\nbar-1)\to -1$,
and since this function is spatially constant, it gives very little 
large-scale power. We therefore retained the radial selection function 
corresponding to the Schecter parameters measured from the data in the
conventional way (de Lapparent {\etal} 1989) in our quoted results above.
However, in light of the sensitivity of the results to slight changes in 
the assumed luminosity function, the leftmost bandpower measurement should be
taken with a liberal dash
of salt. In particular, the suspiciously high value 
seen at $k\sim 0.03/h^{-1}\Mpc$ for the $130h^{-1}\Mpc$
sample in \fig{130fig} may well be due to this effect.
Selection function problems may be present even in the nominally
volume-limited surveys, since Malmquist bias produces a selection
function that is not quite constant near the far edge of the survey volume.

\section{Conclusion}
\label{ConclusionSec}

We have computed the redshift space power spectrum of the UZC
galaxy redshift catalog. 
The results are summarized in Figures~\ref{101fig}-\ref{magFig} 
and are well fit by, \eg, a 
$\Lambda$CDM model with a moderate amount of bias (1.2 to 1.4).

This paper is part of a larger effort to make galaxy redshift survey
analysis more comparable to the state of the art for CMB experiments, 
explicitly calculating the window function for each band power
measurement and including all finite-volume effects in the quoted
error bars. However, it is merely the first step, and much 
work remains to be done:
\begin{enumerate}

\item We measured only the power spectrum in redshift space. 
A more ambitious analysis should extract the additional information
present in clustering anisotropies due to redshift-space distortions.

\item 
We focused only on large scales, optimizing our pair weighting 
for the case of Gaussian fluctuations.
For future work on smaller non-linear scales, 
$k\sim 0.3/h^{-1}\Mpc$,
better results can be obtained by modeling the non-negligible correlations
between different $\k$-modes that have been observed in simulations
(Meiksin \& White 1999; Scoccimarro {\etal} 1999; Hamilton 2000).

\item We measured merely the galaxy power spectrum. A more ambitious 
analysis would attempt to model the biasing issues needed to 
translate this into a measurement of the underlying
matter power spectrum, for instance by combining 
the stochastic biasing formalism 
(Pen 1998; Dekel \& Lahav 1999; Tegmark \& Peebles 1998;
Somerville {\etal} 2000)
with redshift space distortions 
as outlined in Tegmark (1998) and/or by studying higher-order 
moments.

\item Given the sensitivity of the results to errors in 
the radial selection function, it would be worthwhile for a future
analysis to include a likelihood analysis of its shape 
(Binggeli {\etal} 1988; Willmer 1997; Tresse 1999).

\item It is likely that model
testing with future large surveys will require detailed
Monte Carlo simulations to quantify and correct for subtle survey selection
effects (\eg, fiber collisions). Fortunately, the matrix-based
analysis pipeline we have presented lends itself well to this:
once the Fisher matrix has been computed,
the time required to analyze another (Monte Carlo) survey scales merely
as $N_x^2$, not as $N_x^3$.

\item Our clustering analysis was limited to the power spectrum, \ie, to second
moments of the density field. 
Higher-order moments of the density field
are likely to contain a wealth of additional cosmological 
information on small scales 
(see Frieman \& Gazta{\~n}aga 1999 and references therein).

\end{enumerate}
These are major challenges, but the unprecedented quality of
the impending data sets from ongoing redshift surveys such as 2dF and SDSS
provide ample motivation to pursue them.

\bigskip
{\bf Acknowledgements:}
We thank Emilio Falco, Eric Gawiser, Andy Taylor and Jeffrey Willick for useful
discussions, the UZC team for kindly making their data public,
and John Bahcall and Jeffrey Willick for helping to organize a  
summer visit for N.P. to the Institute for Advanced Study during 
which much of this work was carried out.
N.P. and A.J.S.H. both appreciated the hospitality of the IAS.
N.P. was supported by a President's Scholar Grant from Stanford
University.
M.T. was supported by the Alfred P. Sloan Foundation and by 
NASA though grant NAG5-9194 and 
Hubble Fellowship HF-01084.01-96A from STScI (operated by AURA, Inc. 
under NASA contract NAS5-26555). 
A.J.S.H. was supported by NASA grant NAG5-7128.


\bigskip
\bigskip
\bigskip
This paper is available with figures and power spectrum
data from\\ 
{\it h t t p://www.physics.upenn.edu/$\sim$max/uzc.html}


\end{document}